\documentclass[12pt,preprint]{aastex}

\usepackage{epsfig}
\usepackage{natbib}                

        \shorttitle{[L72]~LH~54-425}
        \shortauthors{Williams et al.}

\begin{document}


\title{Dynamical Masses for the Large Magellanic Cloud Massive Binary 
  System [L72] LH~54-425\footnote{Based on observations with the CTIO SMARTS 
    Consortium 1.5-meter telescope and the 2.5-meter DuPont telescope at Las 
    Campanas.}}

\author{S. J. Williams, D. R. Gies, T. J. Henry}
\affil{Center for High Angular Resolution Astronomy and \\
 Department of Physics and Astronomy,\\
 Georgia State University, P. O. Box 4106, Atlanta, GA 30302-4106; \\
 swilliams@chara.gsu.edu, gies@chara.gsu.edu, thenry@chara.gsu.edu}

\author{J. A. Orosz}
\affil{Department of Astronomy, San Diego State University, 5500 Campanile
  Drive, San Diego, CA 92182-1221; orosz@sciences.sdsu.edu}

\author{M. V. McSwain} 
\affil{Department of Physics, Lehigh University, 16 Memorial Drive East,
  Bethlehem, PA 18015; mcswain@lehigh.edu}

\author{T. C. Hillwig}
\affil{Department of Physics and Astronomy, Valparaiso University, Valparaiso,
  IN, 46383; todd.hillwig@valpo.edu}

\author{L. R. Penny}
\affil{Department of Physics and Astronomy, College of Charleston, 101
Science Center, 58 Coming Street, Charleston, SC 29424; pennyl@cofc.edu}

\author{G. Sonneborn}
\affil{NASA Goddard Space Flight Center, Code 681, Greenbelt, MD 20771;
  george.sonneborn@nasa.gov}

\author{R. Iping}
\affil{Department of Physics, Catholic University of America, Washington, DC
  20064; and NASA Goddard Space Flight Center, Code 665, Greenbelt, MD 20771;
  Rosina.C.Iping@nasa.gov }

\author{K. A. van der Hucht}
\affil{SRON Netherlands Institute for Space Research, Sorbonnelaan 2, 3584CA
Utrecht, The Netherlands; Astronomical Institute
Anton Pannekoek, University of Amsterdam, Kruislaan 403, 1097 SJ Amsterdam,
The Netherlands; k.a.van.der.hucht@sron.nl}

\author{L. Kaper}
\affil{Astronomical Institute, University of Amsterdam, Kruislaan 403, 1098
SJ Amsterdam, The Netherlands; lexk@science.uva.nl}

\begin{abstract}

We present results from an optical spectroscopic investigation of the
massive binary system [L72] LH~54-425 in the LH 54 OB association in the 
Large Magellanic Cloud. We
revise the ephemeris of [L72] LH~54-425 and find an orbital period of
2.24741 $\pm$ 0.00004 days. We find spectral types of 
O3 V for the primary and O5 V for the secondary. 
We made a combined solution of the radial velocities and previously published 
$V$-band photometry to determine the inclination for two system configurations,
$i = 52^{+2}_{-3}$ degrees for the configuration of the secondary 
star being more tidally distorted and $i = 55 \pm 1$ degrees for the primary
as the more tidally distorted star. We argue that the latter case is more
probable, and this solution yields masses and radii of 
$M_1 = 47 \pm 2~M_{\odot}$
and $R_1 = 11.4\pm 0.1~R_{\odot}$ for the primary, and
$M_2 = 28 \pm 1~M_{\odot}$ and $R_2 = 8.1 \pm 0.1~R_{\odot}$ for the secondary.
Our analysis places LH~54-425 amongst the most massive stars known. 
Based on the position of the two stars plotted on a theoretical 
HR diagram, we find the age of the system to be $\sim$1.5 Myr.

\end{abstract}


\section{Introduction}

One of the most elusive of important parameters in stellar astrophysics is
the mass of stars. Stars in binary or multiple systems, and specifically
in double-lined spectroscopic binaries, offer us a chance to
explore their masses with one important caveat: we must know the orbital 
inclination of the system. The inclination is the key that allows the 
resolution of the spectroscopic results into individual masses for the 
members of the system.
Individual stellar masses are obtained via the serendipitous orientation of 
the system for eclipses or, at the very least, ellipsoidal light variations. 
Particularly rare are O-type main-sequence systems, owing partly to their 
short lifespans and partly to the steepness of the initial mass function. 
Reported here is the first of several systems to be 
investigated with the goal of determining the masses, radii, and distances to a
number of O-star binaries. Mass data of this kind
are needed to help resolve the mass discrepancy problem
between the evolutionary and spectroscopic mass estimates
for massive stars \citep{rep04,mas05}. 

The earliest type star in the Large Magellanic Cloud LH 54 OB association is
[L72] LH~54-425 \citep{hill94}, also 
known as L54S-4 \citep{oey96a}. Its spectral type was classified by 
\citet{oey98} as O4~III(f*).  
Its binary nature was unknown until \citet{ost02} noted the 
ellipsoidal variations in the $V$-band photometry and obtained six 
spectroscopic observations of the system. The light and radial velocity curves 
from that study were used to estimate
the masses of the two stars at 100 $M_{\odot}$ and 50 $M_{\odot}$. The 
lack of phase coverage for the radial velocity data did not allow for a more
rigorous analysis of physical parameters.

We present here an analysis of the radial velocities from optical spectra
obtained between 2003
and 2006 (\S 2). In \S 3 we discuss how we extracted radial velocities from 
the observations and
solved for the orbital parameters. 
We describe tomographic reconstructions of the individual component spectra
in \S 4. The next section (\S 5) covers our combined light curve and
radial velocity solution, as well as our determination of the inclination
of the system for two system configurations. 
We conclude with a discussion of the
system's evolutionary status, fundamental parameters, and a comparison of
the two model fits to the system in \S 6.


\section{Observations}

The first sets of spectroscopic data were obtained from 2003 through 2006 with 
the Cerro Tololo Inter-American 
Observatory (CTIO) 1.5-m telescope and RC spectrograph operated by the
SMARTS Consortium.
We used grating $\#47$ (831 lines per mm, $8000$~\AA
~blaze wavelength in the first order Littrow configuration) in
second order (SMARTS configuration 47/IIb) together with a
BG39 or CuSO$_4$ order blocking filter.  The detector was a
Loral $1200\times 800$ CCD.  This arrangement produced spectra
covering the range 4058 -- 4732 \AA ~with a resolving power of 
$R=\lambda / \Delta \lambda \approx 3550$.
Exposure times were 1800 seconds, in order to
reach a moderate signal-to-noise ratio 
(S/N $\approx$ 70). Each observation was bracketed
with HeAr comparison spectra for wavelength calibration, and
numerous bias and flat field spectra were obtained each night.
We obtained a total of 48 spectra of the target with this configuration.

In addition to the CTIO data, four observations were obtained with the echelle 
spectrograph on the 2.5-m DuPont telescope at Las Campanas Observatory binned
$2\times 2$ with a resolving power of $\approx$ 27000 (from measurements of arc
spectra). Exposures were combined where there were two, and the spectral
flux was normalized. Orders 
were merged and cosmic rays removed. For continuity of analysis, 
these data were clipped and re-sampled to match the wavelength and lower 
resolution CTIO data.

The resulting spectra were extracted and wavelength calibrated using standard
routines in IRAF\footnote{IRAF is distributed by the
National Optical Astronomy Observatory, which is operated by
the Association of Universities for Research in Astronomy, Inc.,
under cooperative agreement with the National Science Foundation.}. All spectra
were then rectified to a unit continuum via fitting line-free regions and
transformed to a common heliocentric wavelength grid in $\log\lambda$ 
increments.


\section{Radial Velocities and Orbital Elements}

Radial velocities were measured using a template-fitting scheme \citep{gie02}
for the five lines H$\delta$ $\lambda$4101, \ion{He}{2} $\lambda$4200, 
H$\gamma$ $\lambda$4340,
\ion{He}{2} $\lambda$4542, and \ion{He}{2} $\lambda$4686. These were the only 
prominent lines in the spectrum of each star, and no single line showed any 
systematic
shift from the others. The spectra show no evidence for wind effects in the
lines used to derive the orbital solution.
No clear emission features and no asymmetries
of absorption features were seen in the hydrogen lines or the \ion{He}{2} 
$\lambda$4686 line. The basic procedure followed to obtain radial velocities
is outlined in \citet{boy07}. 

The spectrum from HJD 2,452,718.5370 shows well separated
spectral features from each star in the \ion{He}{2} $\lambda$$\lambda$4542, 
4686 lines. These lines were used to derive
matching template spectra for the primary and secondary components. Template 
spectra were obtained from the grid of O-type stellar models from 
\citet{lanz03}, which
are based on the line blanketed, non-LTE, plane-parallel, hydrostatic
atmosphere code $TLUSTY$ and the radiative transfer code $SYNSPEC$ 
\citep{hub88,hub95,hub98}. In order to find matching templates, approximations 
were initially used for the temperatures, gravities, projected rotational 
velocities, and flux contributions from each star. These initial parameters
for model templates were then checked 
after velocity analysis by studying the properties of the tomographically
reconstructed spectra of the components (\S 4).

Further inputs for the template fitting scheme are preliminary estimates of
the velocities for the primary and secondary. The \ion{He}{2} $\lambda$4542 
line for each star was fit using Gaussian functions with the $splot$ routine 
and deblend option in IRAF. 
Radial velocity shifts were then calculated 
for all spectra with well separated lines,
and these were used to obtain initial orbital parameters. The 
preliminary velocity from the initial orbital solution gave a starting point 
to perform a non-linear, least squares fit of the composite
profiles with the template spectra and calculate the shifts for each star. 
The five values for radial velocity from each set of five lines in an 
individual spectrum were then averaged
to obtain a mean value, and the standard deviation of the mean value was 
calculated. Each of these values are listed in Table \ref{rvs} for the primary 
and secondary stars, as well as the orbital phase for each observation and the 
observed minus calculated values for each point. Zero phase is defined by the 
time of inferior conjunction of the primary star, $T_{\rm IC,1}$. 

We made an orbital fit of the radial velocity data using the non-linear,
least-squares fitting program of \citet{mor74}. 
Equal weights were assigned to each data point, because all the spectra were 
comparable in S/N ratio, spectral coverage,
resolving power, and quality of radial velocities derived. 
It should be noted that for massive binaries, discrepancies in
systemic velocities between components may be attributed to differences in
their expanding atmospheres and/or differences in the shapes of template
spectral lines, so we fit systemic velocities for each component. Fits with
non-zero eccentricities were explored, none of which gave as good a fit as a
circular orbital solution. Fits were also made for subsets of measurements near
conjunction (phases near 0.0 and 0.5) and quadrature (phases near 0.25 and 
0.75). These tests revealed that the important spectroscopic elements, such as 
the velocity semi-amplitudes, did not vary by more than 3 $\sigma$ from the 
best fits for any subset.
Our results for the orbit using spectroscopic data only are listed in column 3 
of Table \ref{orbs}. The radial velocity measurements and final orbital
velocity curves (\S 5) are shown in Figure \ref{fig1}.


\section{Tomographic Reconstruction}

We used a Doppler tomography algorithm \citep{bag94} to separate the primary
and secondary spectra of [L72] LH~54-425. This is an iterative method that
uses the 48 observed composite spectra from CTIO, their velocity shifts, and an
assumed monochromatic flux ratio ($F_2$/$F_1$) to derive the individual 
component spectra. We explored a range of flux ratios to arrive at a value
that best matched the line depths in the reconstructions with those in 
model spectra. Figure \ref{fig2} shows the 
reconstructed spectra for the primary and secondary, as well as synthetic
spectra just below each. Few lines are present in our spectra. However, the
relative strength of the \ion{He}{1} $\lambda$4471 in the secondary is the 
major difference between the two spectra and indicates the cooler temperature 
of the secondary. 

These reconstructed spectra were fit with $TLUSTY/SYNSPEC$ model synthesis
spectra (see \S 3). These matches allow for estimates of stellar parameters
listed in Table \ref{tom-params}. 
The width of the \ion{He}{2} $\lambda$4542 line was 
used to estimate the projected rotational velocity $V~\sin~i$ for each 
star by comparing the reconstructed and model profiles for a grid of test
$V \sin i$ values. Matches 
were made to the H$\gamma$ $\lambda$4340 and \ion{He}{1} $\lambda$4471 lines
to obtain the temperature and surface gravity estimates. Finally, we used the 
ratios of the reconstructed to 
synthetic spectral line depths to estimate that the monochromatic 
flux ratio ($F_2$/$F_1$) in our blue spectra is 0.45~$\pm$~0.10.

A visual inspection of our reconstructed spectra 
match best the O3 V example in Figure 6 of \citet{wal02} for the primary and 
the O5 V example in Figure 7 of \citet{wal02}, and these are the spectral 
types we list in Table \ref{tom-params}.
Unfortunately, our wavelength coverage 
does not include the \ion{N}{4} $\lambda$4058 feature suggested in 
\citet{wal02} 
for spectral classification of the hottest O-stars. The \ion{N}{3} 
$\lambda$4634--42 feature is another good diagnostic for classification of 
early O-type stars, but is too weak in our spectra for measurement. 
We can use the equivalent 
width ratio of \ion{He}{1} $\lambda$4471/\ion{He}{2} $\lambda$4542 calibrated 
in \citet{mat88} for each of our reconstructed spectra to obtain a spectral 
class. This measurement results in a spectral class for the primary of O3 and 
of O5.5 for the secondary. We used our best matched effective
temperatures from the $TLUSTY/SYNSPEC$ models combined with the $T_{\rm eff}$
versus spectral type calibration of \citet{mar05} to estimate spectral types of
O3~V for the primary and O5~V for the secondary. These spectral types and
effective temperatures are also consistent with those shown in 
Figure 1 of \citet{mok07} for LMC O-type dwarfs. 


\section{Combined Radial Velocity and Light Curve Solution}

Light curve data were taken from the $V$-band observations listed in Table 1
of \citet{ost02}. New orbital parameters were then found using the 
Eclipsing Light Curve (ELC) code
\citep{oro00}, which fits the radial velocity and light curves 
simultaneously. The resultant light curve is shown in Figure \ref{fig3}, and
the radial velocity curves are shown in Figure \ref{fig1}, along with the
radial velocity measurements from \citet{ost02}, which were not used for
any fit in this paper.
ELC's genetic optimizer was used to explore the parameter space
and was given wide ranges for each value of period, epoch of inferior
conjunction of the primary $T_{\rm IC,1}$, inclination, 
mass ratio, primary velocity semi-amplitude, and Roche-lobe filling factor for 
each star. The Roche-lobe filling factor is defined as the ratio of the 
radius of the star toward the inner Lagrangian Point ($L_1$) to the distance 
to $L_1$ from the center of the star, $f \equiv x_{\rm point}/x_{L1}$ 
\citep{oro00}; our fits show that our stars do not fill their Roche lobe
(see Table \ref{ELC-params}). 
A by-product of the exploration of parameter space by ELC
was a determination of systemic velocities for each component.

The temperatures of each star were set and not fit because these were well 
constrained by analysis of the reconstructed spectra discussed in \S 4. 
As with the fits of only radial velocities discussed in \S 3, circular orbital
solutions proved best. 

To estimate the 1, 2, and 3 $\sigma$ uncertainties associated with the fitted
parameters, we followed the method discussed in \citet{oro02}.
The values of the seven fitted parameters (plus the two systemic velocities)
listed above were varied in the calculation of
$\sim$3.6$\times$10$^{6}$ light and radial velocity curves. 
This procedure resulted in thorough sampling near the $\chi^{2}_{min}$ point,
making the multi-dimensional surface well populated. 
This allowed for projection of
the $\chi^2$ surface as a function of each parameter or fitted value of 
interest. The
lowest $\chi^2$ is then found for each parameter of interest,
and the 1, 2, and 3 $\sigma$ uncertainties may be estimated by the regions 
where $\chi^2 \le \chi^{2}_{min}$ +1, +4, and +9, respectively. The 
uncertainties listed in the tables are 1 $\sigma$ uncertainties.

Special attention was paid to the inclination of the system. The well-sampled
light curve from \citet{ost02} allows for an exploration of the important 
parameters that cause the ellipsoidal variations in this system: 
inclination and Roche-lobe
filling factors for each star. Ellipsoidal light variations occur in systems
where one or both of the components are tidally distorted due 
to the proximity of its companion. At different orbital phases,
the observer sees varying amounts of light from the system, based upon the 
interplay between the amount of tidal distortion, the observed cross-section
of the two stars, and the inclination of the system. 
As was expected, we found that for lower inclinations, i.e., viewing
the system more face-on, the sizes and tidal distortions of the stars 
needed to become larger in order to match the 
modulation of the light curve. Thus, restricting the stars to radii within 
their Roche lobes ultimately limits the lowest acceptable inclination,
while the lack of observed eclipses establishes the maximum inclination.

Again, using the genetic algorithm in ELC, we
explored the range of inclination values suggested in \citet{ost02} and allowed
the fill factors for each star to vary over a wide range while keeping the
other parameters fixed. The results of this
analysis are plotted as contours of 1, 2, and 3 $\sigma$ confidence intervals 
in Figures \ref{fig4} and \ref{fig5}. 
In the best overall fit, the estimated filling factors are
larger for the secondary, so most of the ellipsoidal variation is due to the
tidal distortion of the secondary (Fig. \ref{fig4}). 
It is interesting to note that the 
contours suggest a rather steep surface toward higher inclinations, indicating
a hard upper limit for the inclination of 56$^{\circ}$ at the 3 $\sigma$ level.
Lower inclinations are not as probable,
and fill factors for the secondary get too large to match the light curve
as the secondary star approaches a fill factor of 1.0, thus completely filling
its Roche-lobe. These characteristics
help us to constrain the inclination of the system to be $i = 52^{+2}_{-3}$
degrees. 
The corresponding masses and other parameters are listed in Table 
\ref{ELC-params}. Included
are values for $V_{\rm sync}~\sin~i$, the projected rotational velocity 
assuming synchronous rotation, $R_{\rm eff}$, the effective radius for a 
sphere of the same volume, and the sizes of the stars along different
directions. 

The orbital parameters for the best fit are listed in Table \ref{orbs}.
The spectroscopic solution and the combined solution from ELC match very well.
It should be noted that the parameters in Table \ref{ELC-params} 
with asymmetric uncertainties are a result of the asymmetry
in the uncertainties for the inclination. 

We also explored the possibility of the primary being the more 
tidally distorted star. 
To do this we ran the simulations again, keeping the same parameters from 
Table \ref{orbs} and varying only the filling factors and inclination. 
Thus, the
radial velocity curve does not change and the light curve is fit to roughly 
the same degree of accuracy as before, so the fits in Figures \ref{fig1} and
\ref{fig3} need not be changed.
The contour plot for this scenario,
again with confidence intervals of 1, 2, and 3 $\sigma$, is shown in Figure
\ref{fig5}. The fit gives an inclination of $i = 55 \pm 1$ degrees. This 
corresponds to slightly different values of masses and radii, listed alongside
the overall best fit in Table \ref{ELC-params}. The small range in inclination 
leads to a small range in the masses and radii of the components of the system.
This restricted fit has a reduced $\chi^2$ of 2.69 versus 2.54 for the case 
where the secondary is more tidally distorted. 

A contact binary is another possible configuration for the system. We did not
explore this possibility because in a contact system, the sum of the values of 
$V \sin i$ and the sum of the velocity semi-amplitudes would be comparable. For
our fits of [L72] LH~54-425 (see Tables \ref{orbs} and \ref{tom-params}), 
the sums of these values differ by at least 300 km s$^{-1}$. 


\section{Discussion and Conclusions}

Which of our two models is most consistent with the available
data? The answer to this question lies in comparing the 
consequences of each solution.

To begin the comparison, we used the temperatures and radii output from
ELC to obtain an estimate of the monochromatic flux ratio $F_2$/$F_1$.
The case where the secondary is more tidally distorted
yields a model flux ratio of 0.70~$^{+0.05}_{-0.01}$, while the primary as the 
more tidally distorted star gives a flux ratio of 0.46~$\pm$~0.02, in agreement
with
the value of 0.45~$\pm$~0.05 derived from the tomographic reconstructions.

Theory suggests \citep{eks08} that for massive stars, the main sequence radius 
scales as
$R \sim M^{0.6}$. Applying this as another consistency check, we can substitute
velocity semi-amplitudes to estimate the expected ratio of radii, 
$R_2/R_1\sim(K_1/K_2)^{0.6}$ which, from our values 
in Table \ref{orbs} is 0.707 $\pm$ 0.10. The 
ratio of effective radii estimated from the 
output of  ELC (Table \ref{ELC-params}) 
is $\sim0.88^{+0.03}_{-0.05}$ for the fit with the secondary as the more
tidally distorted and $\sim0.71\pm0.010$ 
for the primary being more tidally distorted. Once again the fit with the
primary as the more tidally distorted matches theory more closely.

Ignoring the tidal distortions, we can estimate the luminosity
by applying the Stefan-Boltzmann equation $L=4\pi R^2 \sigma T^{4}_{\rm eff}$, 
and combining this with our temperature measurements, 
place each component in a theoretical
H-R diagram. This is presented in Figure \ref{fig6}, along with evolutionary
tracks from \citet{scha93} for stars of varying mass with a metallicity of
$Z = 0.008$ which is appropriate for the LMC. The 1 
$\sigma$ uncertainty regions are consistent with an age of $\sim$1.5 Myr for 
the model consisting of the primary as the more tidally distorted star, and
both stars appear to be co-evolutionary. 
The tidally distorted secondary fit appears a bit older, 
roughly matching the 2 Myr isochrone. This fit also indicates that the 
secondary is more evolved than the primary and is overluminous for
its derived mass. Therefore, it seems the fit with the primary being more
tidally distorted is more consistent with evolutionary tracks. 
These approximate ages are also consistent within uncertainties 
with the estimate 
from \citet{oey98} of $\sim$2-3 Myr for the LH 54 OB association. 
According to Table 1 of \citet{scha93}, even the oldest age of $\sim$2 Myr
puts [L72] LH~54-425 in the part of its life when it is still burning 
hydrogen, and it will continue to do so for another $\sim$1.5--2 Myr. 

For the next comparison, we used the luminosities calculated above to obtain a 
bolometric absolute magnitude, assuming a solar bolometric absolute magnitude 
of 4.74 mag \citep{cox00}. Next, we applied a bolometric correction for each 
star based on our derived spectral types from \citet{mar05} to obtain absolute 
$V$ magnitudes. We then combined these magnitudes to obtain an absolute $V$ 
magnitude for the system. Finally, we arrived at a distance
modulus for the system by using the maximum magnitude of the system from the
photometric data and $E(B-V)$ = 0.10 mag \citep{oey98}. The resulting 
distance modulus from the fit with the more tidally distorted secondary is 
18.65 mag while the distance modulus for the tidally distorted primary
fit is 18.55 mag. Both of these numbers are consistent with the HST Key 
Project distance modulus to the LMC of 18.50 $\pm$ 0.10 mag \citep{fre01},
with the tidally distorted primary fit being slightly better. 
Note that each of the contour plots in Figures \ref{fig4} and \ref{fig5} 
seem to have 
a best fit valley in the $\chi^2$ as a function of inclination and filling
factor. As one travels along this valley to higher inclinations, the
dimensions of the system get smaller, including the sizes of the two stars.
This makes the stars intrinsically fainter, and they must lie at a closer 
distance to match the observed $V$ photometry. In contrast, the lower 
inclinations mean the stars become bigger and lie at greater distances.
The overall agreement in distance modulus between the model results and
those for the LMC indicate that the final fill out estimates are reliable. 

The two models 
are most different in the $V_{\rm sync}~\sin~i$ values output from ELC. 
The observed $V~\sin~i$ values from the reconstructed spectra are 
197~$\pm$~5 km s$^{-1}$ for the primary and 182~$\pm$~8 km s$^{-1}$ 
for the secondary. The ELC fit for the more 
distorted secondary gives values of 197~$\pm$~5 km s$^{-1}$ and 173~$\pm$~7 
km s$^{-1}$ for the
primary and secondary, respectively, while the ELC fit for the more distorted
primary gives 209~$\pm$~5 km s$^{-1}$ and 148~$\pm$~6 km s$^{-1}$. 
Both cases agree within uncertainties for the primary's line broadening, but
the predicted $V \sin i$ for the secondary is too small for the case where
the primary is the more tidally distorted star.
However, due to the youth of the system, our assumption of
synchronous rotation may not be appropriate. There may not have yet been enough
time for tidal forces to synchronize the spin with the orbit. More ELC fits
were performed assuming a faster rotation for the secondary and these show 
that the same physical parameters are obtained for a non-synchronous rotation 
scenario.

\begin{figure}
\begin{center}
{\includegraphics[angle=90,height=12cm]{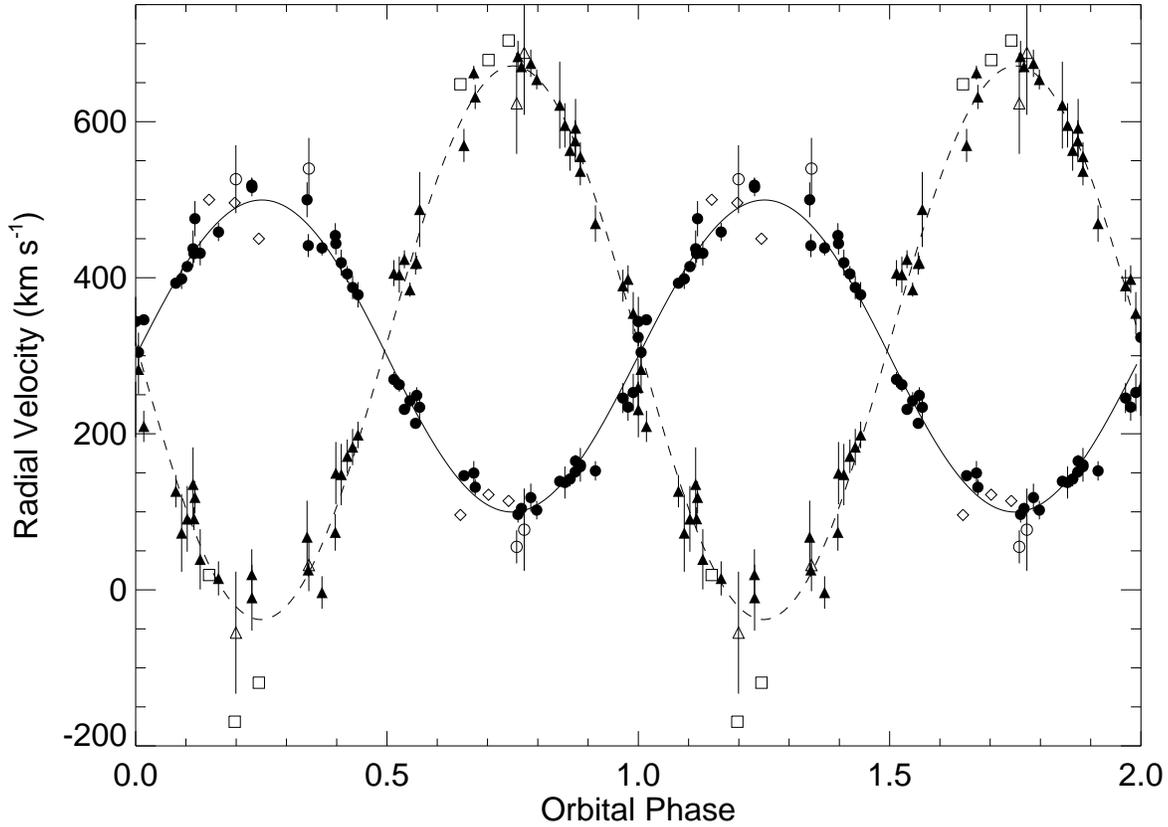}}
\end{center}
\caption{Radial velocity curves for [L72] LH~54-425. Primary radial velocities
  are shown by filled dots and secondary radial velocities are represented by 
  filled triangles with associated uncertainties shown as line segments for 
  both. The open circles and triangles represent the echelle
  data from Las Campanas. The open diamonds and squares are the radial velocity
  measurements from \citet{ost02} for the primary and secondary, respectively.
  The Ostrov data were not used in model fits.
  The solid line is the best combined fit solution for the primary, and the 
  dotted line is the same for the secondary.}
\label{fig1}
\end{figure}

\begin{figure}
\begin{center}
{\includegraphics[angle=90,height=12cm]{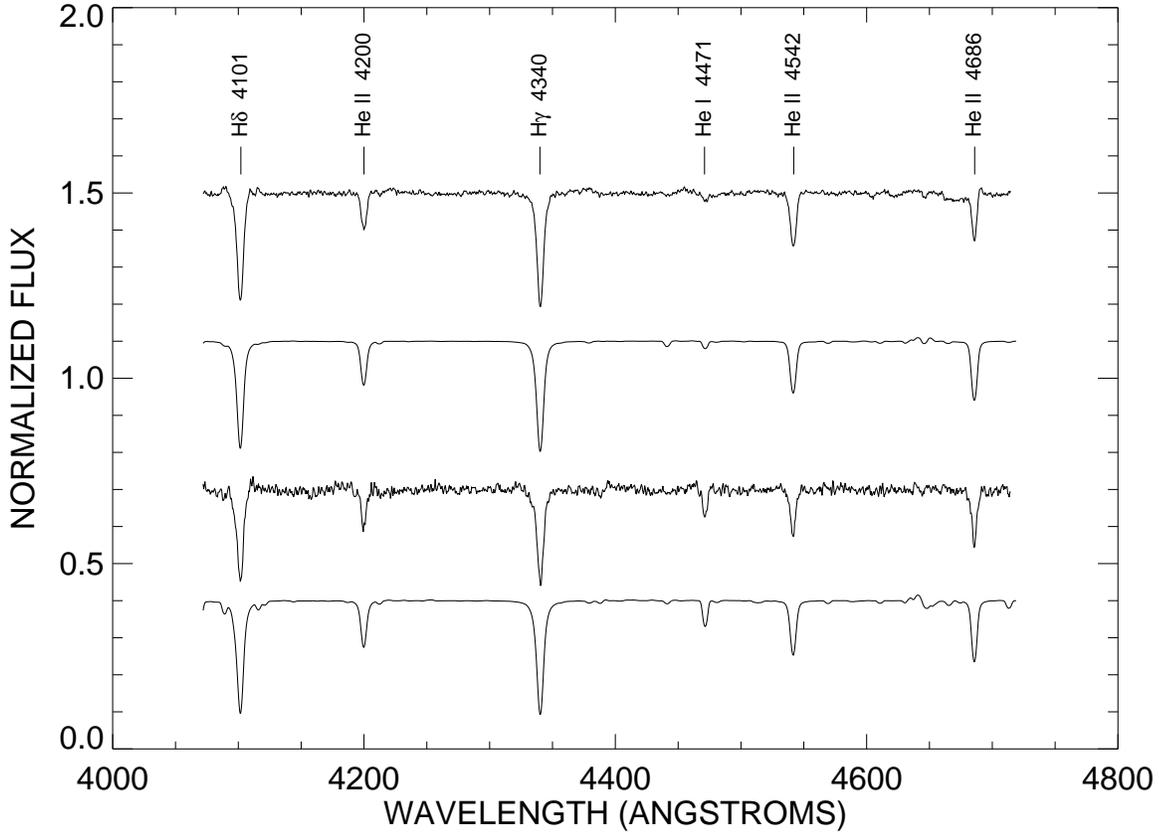}}
\end{center}
\caption{Tomographic reconstructions of the components of [L72] LH~54-425
  based on 48 spectra obtained from 2003 through 2006 at CTIO. Plotted from top
  to bottom are line identifications with vertical tick marks, the primary 
  spectrum, the model primary spectrum \citep{lanz03}, the secondary spectrum,
  and the model secondary spectrum. The stellar parameters for the model
  spectra are given in Table \ref{tom-params}.}
\label{fig2}
\end{figure}

\begin{figure}
\begin{center}
{\includegraphics[angle=90,height=12cm]{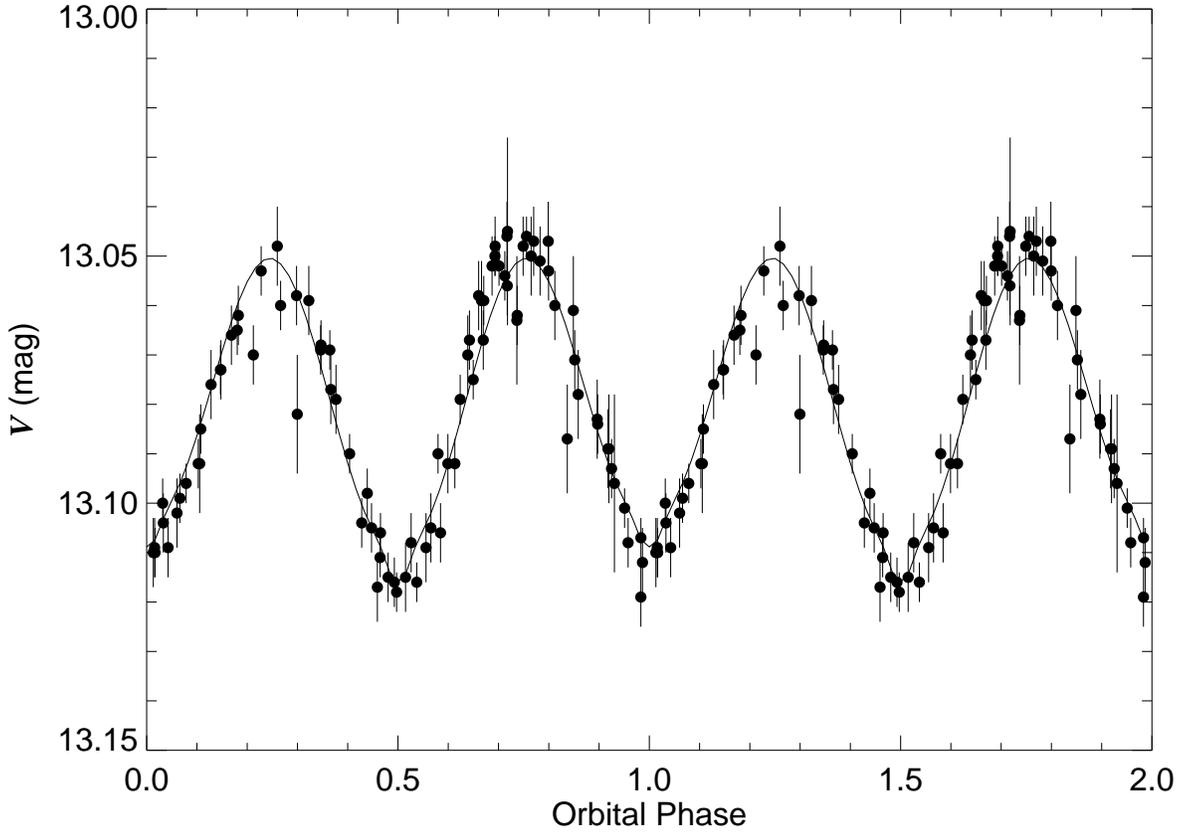}}
\end{center}
\caption{$V$-band light curve for [L72] LH~54-425. These data were taken from 
  \citet{ost02} and are presented here in phase according to our 
  best combined solution. The model is the solid line and the data are 
  represented by filled dots with $V$ uncertainties shown by line segments.
  Phase zero corresponds to inferior conjunction of the primary star
  (which differs by 0.5 phase from that adopted by \citealt{ost02}).}
\label{fig3}
\end{figure}

\begin{figure}
\begin{center}
{\includegraphics[angle=90,height=12cm]{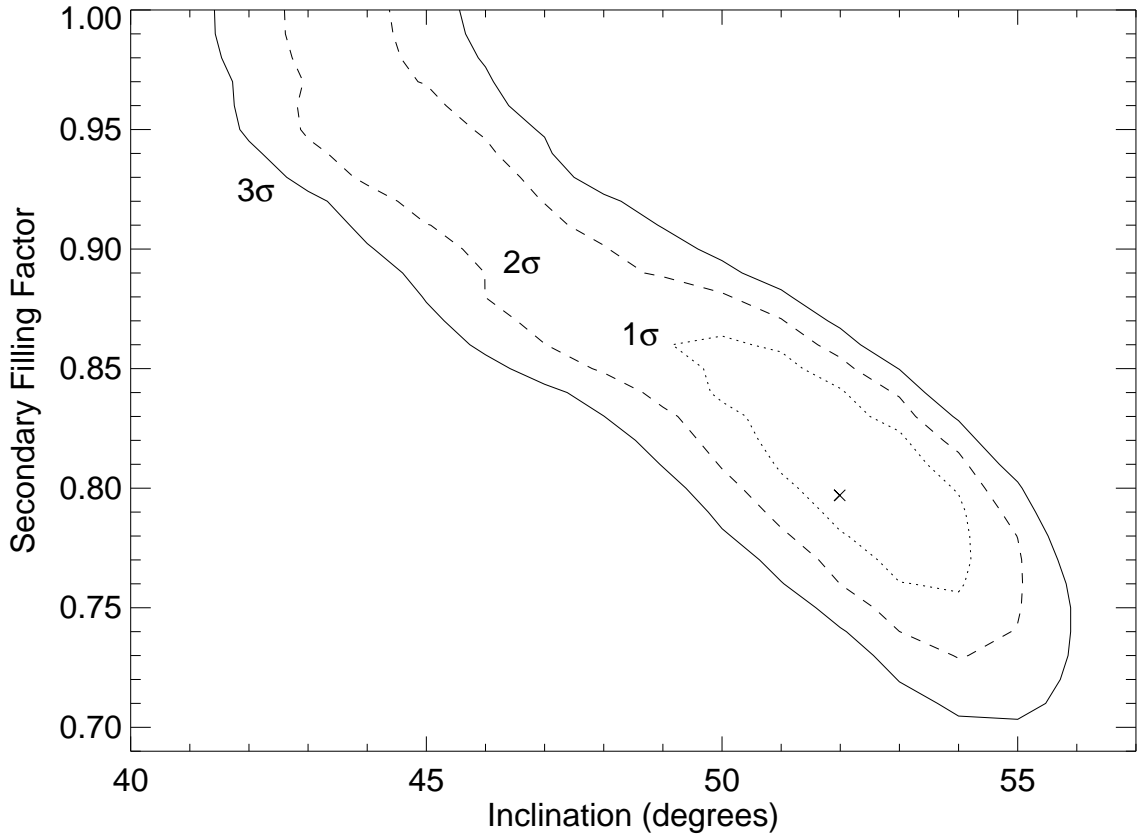}}
\end{center}
\caption{$\chi^2$ surface contours of the residuals from the combined velocity
  and light curve data as a function of orbital inclination and secondary 
  Roche-lobe filling factor of the best fit solution. The best fit position is 
  represented by the ``x'' inside the 1-$\sigma$ contour. The range of filling 
  factor is quite large, while the inclination is reasonably well 
  constrained.}
\label{fig4}
\end{figure}

\begin{figure}
\begin{center}
{\includegraphics[angle=90,height=12cm]{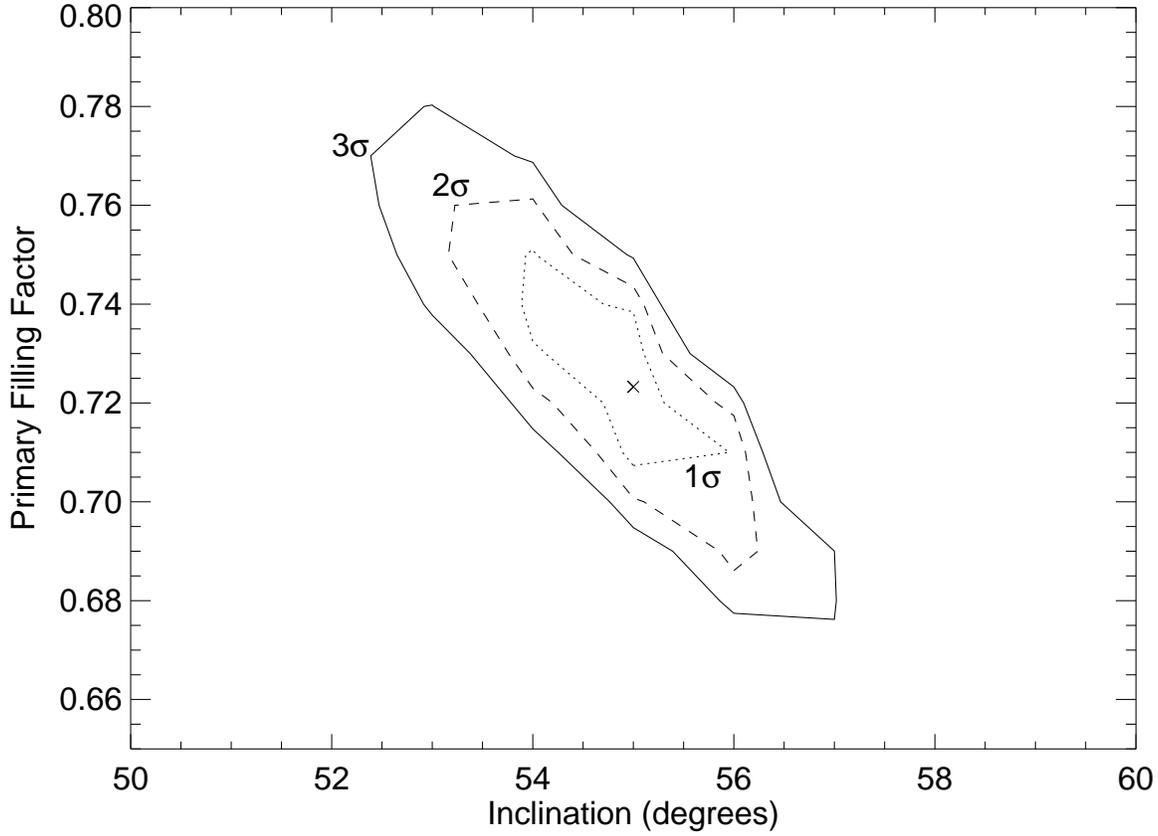}}
\end{center}
\caption{$\chi^2$ surface contours of the combined fit residuals as a function
  of the orbital inclination and Roche-lobe filling factor of the primary 
  star for the case where the primary is the more tidally distorted star. 
  The best fit position 
  is represented by the ``x'' inside the 1-$\sigma$ contour. The parameter 
  space is more confined than for the best fit, and the contours are smaller 
  around the best inclination for this configuration.}
\label{fig5}
\end{figure}

\begin{figure}
\begin{center}
{\includegraphics[angle=90,height=12cm]{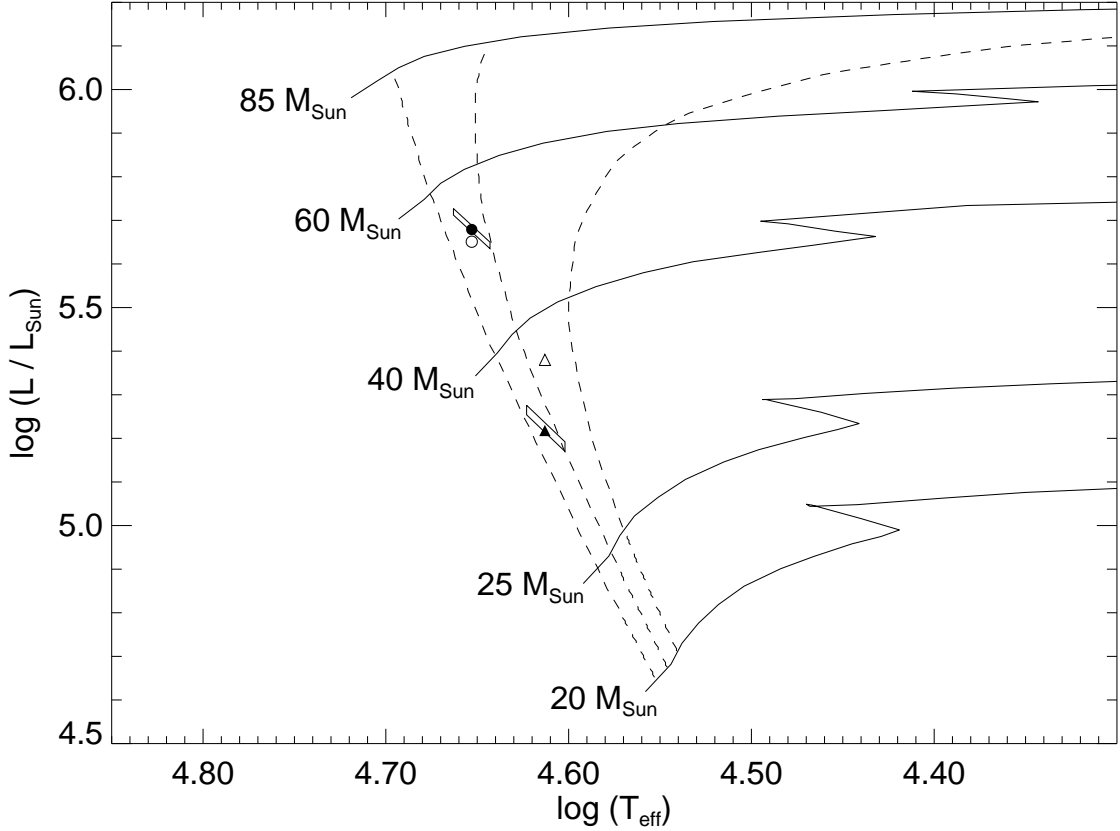}}
\end{center}
\caption{An H-R diagram showing the location of the primary star 
  ($filled~circle$) and secondary star ($filled~triangle$) of [L72] LH~54-425 
  for our fit where the primary is the more tidally distorted star. The open 
  symbols represent the fit where the secondary is the more tidally distorted 
  star. Also plotted are
  evolutionary tracks for stars of various masses from \citet{scha93} for an 
  LMC metallicity. The boxes
  around the filled data points correspond to the 1-$\sigma$ uncertainties in 
  the derived values of $T_{\rm eff}$ and $R$ from the  
  fits where the primary star is the more tidally distorted star. 
  The vertical dashed lines correspond to isochrones from 
  \citet{lej01}, also for an LMC metallicity, of 1, $\sim$2 and $\sim$3.2 Myr 
  going from left to right. The positions of the two stars are consistent
  with an age of $\sim$1.5 Myr.}
\label{fig6}
\end{figure}

Given the above tests, we think the fit with the primary as the more tidally 
distorted star is the more likely configuration. For this case, the masses 
and radii are
$M_1 = 47 \pm 2~M_{\odot}$ and $R_1 = 11.4~\pm~0.1~R_{\odot}$ for the primary
and $M_2 = 28 \pm 1~M_{\odot}$ and $R_2 = 8.1~\pm~0.1~R_{\odot}$ for secondary.
These masses are significantly less than the 100 $M_{\odot}$ and
50 $M_{\odot}$ values obtained by \citet{ost02}. Part of the reason lies in the
data set we obtained for radial velocities of the system. As is shown in
Figure \ref{fig1}, two secondary velocity values derived by \citet{ost02} near 
phase 0.25 are the ones that drove the estimate for the velocity 
semi-amplitude up,
and consequently the mass ratio down. This, in addition to the lower 
inclination adopted by \citet{ost02}, leads to the differences in masses 
between our two analyses. Our temperatures are also slightly lower than those 
given in \citet[][ see his Table 3]{ost02}, due to his use of a different, 
older temperature versus spectral-type calibration \citep{all82}. We can
compare our primary star with the primary in R136-38 \citep{mas02}, which is
also an O3 V. \citet{mas02} derive a mass for their O3 V star of 
$56.9 \pm 0.6~M_{\odot}$, in closer agreement with our value of 47~$M_{\odot}$
than that of 100~$M_{\odot}$ found by \citet{ost02}. Our derived
radius for the primary in LH 54-425 is slightly larger than that for the 
primary in R136-38
of 9.3 $R_{\odot}$ \citep{mas02}. This may be
due to the extreme youth of R136-38, which appears to be
near the zero-age main sequence \citep[see Fig. 6 in ][]{mas02}, 
while the position of [L72] LH~54-425 in the H-R diagram (our Fig. \ref{fig6}) 
indicates the primary is slightly evolved. These numbers are
consistent with the fact that the LH 54 OB association, at $\sim$2--3 Myr
\citep{oey98}, is slightly older than the R136 cluster with an age of 
$\sim$1--2 Myr \citep{mas98}.


\acknowledgments

We gratefully acknowledge the referee for comments and suggestions 
that improved the quality and organization of the paper. We also
gratefully acknowledge the 2.5-meter DuPont data and thoughtful
comments and suggestions generously offered by Alceste Bonanos.
This material is based upon work supported by the National Science
Foundation under Grant No.~AST-0506573 and AST-0606861.
We gratefully acknowledge support from the GSU College
of Arts and Sciences and from the Research Program Enhancement
fund of the Board of Regents of the University System of Georgia,
administered through the GSU Office of the Vice President for Research. 
We also wish to thank Fred Walter and the SMARTS Consortium for a generous 
allocation of time to study this system with the CTIO 1.5-m telescope. 


\bibliographystyle{apj}

\bibliography{apj-jour,paper}


\newpage

\begin{deluxetable}{cccccccc}
\tabletypesize{\scriptsize}
\tablewidth{0pt}
\tablecaption{LH~54--425 Radial Velocity Measurements\label{rvs}}
\tablehead{
\colhead{Date}          &
\colhead{Orbital}       &
\colhead{$V_1$}         &
\colhead{$\sigma_{1}$}  &
\colhead{$(O-C)_1$}     &
\colhead{$V_2$}         &
\colhead{$\sigma_{2}$}  &
\colhead{$(O-C)_2$}     \\
\colhead{(HJD$-$2,400,000)}        &
\colhead{Phase}  &
\colhead{(km s$^{-1}$)} &
\colhead{(km s$^{-1}$)} &
\colhead{(km s$^{-1}$)} &
\colhead{(km s$^{-1}$)} &
\colhead{(km s$^{-1}$)} &
\colhead{(km s$^{-1}$)} }
\startdata
 52714.5496 & 0.080 & 393.0    & \phn6.7 & \phn--4.1    & 126.4     & \phn20.7  & --16.2      \\
 52714.5756 & 0.092 & 398.5    &   12.9  & --11.1       & \phn72.8  & \phn49.6  & --47.5      \\
 52714.6009 & 0.103 & 414.3    & \phn4.9 & \phn--6.7    & \phn90.9  & \phn42.1  & \phn--9.2   \\
 52714.6312 & 0.116 & 430.9    &   11.7  & \phn--3.1    & \phn91.2  & \phn14.6  & \phn14.0    \\
 52714.6586 & 0.129 & 431.4    &   15.6  & --13.7       & \phn39.1  & \phn38.6  & --18.4      \\
 52715.5251 & 0.514 & 269.6    & \phn9.7 & --11.8       & 405.9     & \phn16.7  & \phn57.1    \\
 52715.5487 & 0.525 & 263.0    & \phn9.1 & \phn--4.8    & 404.0     & \phn23.1  & \phn31.4    \\
 52715.5725 & 0.535 & 231.3    & \phn5.7 & --23.8       & 423.6     & \phn11.5  & \phn28.5    \\
 52715.5971 & 0.546 & 242.3    &   11.1  & \phn\phn 0.9 & 384.8     & \phn\phn9.1 & --34.4      \\
 52715.6216 & 0.557 & 213.4    & \phn5.0 & --14.5       & 418.4     & \phn14.3  & --24.4      \\
 52716.5492 & 0.970 & 245.8    &   19.2  & --16.0       & 389.8     & \phn20.5  & \phn\phn4.4 \\
 52716.5722 & 0.980 & 234.2    &   17.4  & --40.4       & 398.1     & \phn17.8  & \phn35.6    \\
 52716.5952 & 0.990 & 252.9    &   24.1  & --34.6       & 354.5     & \phn27.1  & \phn15.3    \\
 52716.6182 & 0.000 & 343.9    &   31.6  & \phn43.4     & 231.1     & \phn35.6  & --84.7      \\
 52717.5139 & 0.399 & 443.8    &   14.0  & \phn23.5     & 149.6     & \phn39.9  & \phn43.9    \\
 52717.5370 & 0.409 & 419.3    &   16.6  & \phn10.1     & 147.8     & \phn39.3  & \phn22.5    \\
 52717.5642 & 0.421 & 404.9    &   10.2  & \phn\phn8.8  & 171.2     & \phn21.5  & \phn22.8    \\
 52717.5878 & 0.432 & 387.5    &   15.0  & \phn\phn3.5  & 183.0     & \phn23.3  & \phn13.3    \\
 52717.6122 & 0.443 & 378.2    &   16.3  & \phn\phn6.8  & 198.6     & \phn17.0  & \phn\phn6.8 \\
 52718.5139 & 0.844 & 139.1    & \phn6.4 & \phn\phn5.6  & 621.2     & \phn55.7  & \phn\phn7.8 \\
 52718.5370 & 0.854 & 137.6    &   20.6  & \phn--3.3    & 595.3     & \phn28.4  & \phn--5.0   \\
 52718.5600 & 0.864 & 142.0    & \phn9.3 & \phn--7.0    & 563.3     & \phn26.1  & --22.8      \\
 52718.5831 & 0.875 & 151.1    &   10.2  & \phn--6.6    & 575.4     & \phn26.5  & \phn\phn4.6 \\
 52718.6060 & 0.885 & 157.9    & \phn7.4 & \phn--9.2    & 536.5     & \phn18.1  & --17.8      \\
 53017.5789 & 0.915 & 152.5    &   12.4  & --44.9       & 469.5     & \phn23.3  & --31.0      \\
 53017.7698 & 0.000 & 323.6    &   25.6  & \phn23.8     & 259.5     & \phn36.7  & --57.5      \\
 53018.5366 & 0.341 & 499.9    &   22.4  & \phn30.8     & \phn67.6  & \phn46.7  & \phn48.4    \\
 53019.5645 & 0.798 & 102.2    &   11.9  & \phn--6.7    & 654.1     & \phn12.7  & \phn--1.9   \\
 53019.7591 & 0.785 & 160.2    &   21.6  & \phn--6.8    & 555.3     & \phn18.5  & \phn\phn0.8 \\
 53020.5382 & 0.232 & 515.8    &   11.4  & \phn17.4     & --10.0    & \phn42.2  & \phn24.9    \\
 53021.5305 & 0.673 & 149.7    &   15.4  & \phn28.0     & 662.8     & \phn\phn8.9 & \phn32.1    \\
 53021.7435 & 0.768 & 104.3    & \phn6.8 & \phn\phn3.2  & 670.6     & \phn\phn3.7 & \phn\phn1.5 \\
 53022.5223 & 0.114 & 437.1    &   23.5  & \phn\phn5.0  & 135.2     & \phn47.4  & \phn54.7    \\
 53023.5216 & 0.559 & 249.2    &   10.4  & \phn23.7     & 419.2     & \phn\phn8.9 & --27.8      \\
 53027.5315 & 0.344 & 441.2    &   14.8  & --26.2       & \phn25.6  & \phn\phn8.6 & \phn\phn3.5 \\
 53028.5271 & 0.786 & 118.3    &   17.8  & \phn13.2     & 674.8     & \phn17.5  & \phn12.3    \\
 53028.7270 & 0.875 & 165.0    & \phn8.0 & \phn\phn6.9  & 591.8     & \phn37.4  & \phn21.7    \\
 53029.5269 & 0.231 & 518.6    & \phn9.8 & \phn20.2     & \phn19.6  & \phn32.4  & \phn54.4    \\
 53030.5260 & 0.676 & 131.5    &   10.3  & \phn11.4     & 631.7     & \phn15.5  & \phn--1.9   \\
 53030.7169 & 0.761 & \phn96.7 &   10.5  & \phn--3.6    & 683.5     & \phn20.2  & \phn13.2    \\
 53031.5200 & 0.118 & 475.7    &   22.6  & \phn40.1     & 118.6     & \phn20.8  & \phn44.2    \\
 53032.5246 & 0.565 & 234.0    &   18.9  & \phn15.5     & 487.5     & \phn48.1  & \phn28.3    \\
 53032.7225 & 0.653 & 146.3    & \phn3.9 & \phn12.1     & 569.6     & \phn21.2  & --38.7      \\
 53033.5391 & 0.017 & 346.1    & \phn6.3 & \phn25.4     & 209.5     & \phn20.0  & --69.8      \\
 53710.7543\tablenotemark{a} & 0.347 & 539.9    &   39.3  & \phn75.4     & \phn31.7  & \phn33.4  & \phn\phn5.4 \\
 53711.7167\tablenotemark{a} & 0.776 & \phn77.1 &   52.7  & --29.5       & 688.2     &  \phn79.4   & \phn23..6   \\
 53748.5493 & 0.165 & 458.6    &   11.7  & --13.4       & \phn14.7  & \phn21.9  & \phn\phn4.2 \\
 53752.6857 & 0.006 & 304.4    &   25.3  & \phn--2.3    & 282.6     & \phn31.5  & --22.0      \\
 53753.5655 & 0.397 & 454.1    &   15.9  & \phn32.0     & \phn73.8  & \phn23.6  & --28.8      \\
 53755.7542 & 0.371 & 438.2    & \phn9.6 & \phn--8.5    & \phn--3.4 & \phn20.9  & --62.5      \\
 53765.6205\tablenotemark{a} & 0.761 & \phn55.1 &   21.2  & --49.6       & 623.4     &  \phn64.7   & --44.4      \\
 53766.6131\tablenotemark{a} & 0.203 & 526.3    &   43.6  & \phn40.6     & --54.9    &  \phn78.2   & --37.6      \\
\enddata

\tablenotetext{a}{Data from the echelle spectrograph on the 2.5 m DuPont 
  telescope at Las Campanas.}

\end{deluxetable}


\begin{deluxetable}{lcc}
\tablewidth{0pc}
\tablecaption{Circular Orbital Solutions for [L72]~LH 54-425\label{orbs}}
\tablehead{
  \colhead{Element}    &
  \colhead{Combined Solution} &
  \colhead{Spectroscopic Solution}}
\startdata
$P$~(days)\dotfill                       & \phn\phn\phn\phn2.24741 $\pm$ 0.00004 & \phn\phn\phn\phn\phn2.24746 $\pm$ 0.00010\\
$T_{\rm IC,1}$ (HJD--2,400,000)\dotfill  & 53029.007 $\pm$ 0.003 & \phn53029.016 $\pm$ 0.007 \\
$K_1$ (km s$^{-1}$)\dotfill              & \phn\phn201.6 $\pm$ 3.8 & \phn\phn\phn210.8 $\pm$ 3.0 \\
$K_2$ (km s$^{-1}$)\dotfill              & \phn\phn359.1 $\pm$ 4.5 & \phn\phn\phn350.8 $\pm$ 5.2 \\
$\gamma_1$ (km s$^{-1}$)\dotfill         & \phn\phn299.7 $\pm$ 0.6 & \phn\phn\phn303.1 $\pm$ 2.0 \\
$\gamma_2$ (km s$^{-1}$)\dotfill         & \phn\phn316.7 $\pm$ 2.2 & \phn\phn\phn317.7 $\pm$ 3.3 \\
rms~(Primary)~(km s$^{-1}$)\dotfill      & \phn\phn19.4 & \phn\phn\phn21.9 \\
rms~(Secondary)~(km s$^{-1}$)\dotfill    & \phn\phn33.2 & \phn\phn\phn32.8 \\
rms~(Photometry)~(mag)\dotfill           & \phn\phn0.007 & \phn\phn\phn\nodata
\enddata
\end{deluxetable}

\begin{deluxetable}{lcc}
\tablewidth{0pc}
\tablecaption{Tomographic Spectral Reconstruction Parameters for [L72]~LH~54-425
              \label{tom-params}}
\tablehead{
  \colhead{Parameter}          &
  \colhead{Primary}            &
  \colhead{Secondary}}
\startdata
Spectral Type                  & O3~V                  & O5~V\\
$T_{\rm eff}$ (kK)             & \phn45 $\pm$ 1        & \phn41 $\pm$ 1\\ 
log $g$ (cgs)                  & \phn\phn4.0 $\pm$ 0.2 & \phn\phn4.0 $\pm$ 0.2\\
$V~\sin~i$ (km s$^{-1}$)       & 197 $\pm$ 5           & 182 $\pm$ 8\\
$F_{\rm 2} / F_{\rm 1}$ (blue) & \multicolumn{2}{c}{0.45 $\pm$ 0.05}\\
\enddata
\end{deluxetable}

\begin{deluxetable}{lcccc}
\tablewidth{0pc}
\tablecaption{ELC Model Parameters for [L72]~LH 54-425\label{ELC-params}}
\tablehead{
  \colhead{Parameter}  &
  \multicolumn{2}{c}{Distorted Secondary}    &
  \multicolumn{2}{c}{Distorted Primary}  \\
  \colhead{\nodata}    &
  \colhead{Primary}    &
  \colhead{Secondary}  &
  \colhead{Primary}    &
  \colhead{Secondary}}
\startdata
Inclination (deg)         & \multicolumn{2}{c}{52 $^{+2}_{-3}$} & \multicolumn{2}{c}{55 $\pm$ 1}\\
$M$ ($M_{\odot}$)         & 53 $^{+7}_{-4}$ & 32 $^{+4}_{-2}$ & ~~47 $\pm$ 2 & ~~28 $\pm$ 1\\
$R_{\rm eff}$ ($R_{\odot}$) & 11.0 $^{+0.7}_{-0.3}$ & \phn9.7 $^{+1.0}_{-0.2}$ & ~~11.4 $\pm$ 0.1 & \phn~~8.1 $\pm$ 0.1\\
$R_{\rm pole}$\tablenotemark{a} ($R_{\odot}$) & 10.7 $^{+0.3}_{-0.2}$ & \phn9.2 $^{+0.3}_{-0.2}$ & ~~10.9 $\pm$ 0.1 & \phn~~7.8 $\pm$ 0.1\\
$R_{\rm point}$\tablenotemark{b} ($R_{\odot}$) & 11.7 $^{+0.4}_{-0.2}$ & 11.1 $^{+0.4}_{-0.2}$ & ~~12.3 $\pm$ 0.2 & \phn~~8.6 $\pm$ 0.1\\
$V_{\rm sync}~\sin~i$ (km s$^{-1}$)  & 197 $\pm$ 5 & 173 $\pm$ 7 & 209 $\pm$ 5 & 148 $\pm$ 6\\
log $g$ (cgs)             & \phn4.08$^{+0.01}_{-0.01}$ & \phn3.94$^{+0.02}_{-0.01}$& \phn\phn4.00 $\pm$ 0.02 & \phn\phn4.07 $\pm$ 0.01\\
Filling Factor            & \phn0.66 $^{+0.04}_{-0.02}$ & \phn0.80 $^{+0.08}_{-0.02}$ & \phn0.72 $^{+0.03}_{-0.02}$ & \phn0.64 $^{+0.03}_{-0.02}$\\
$a_{\rm tot}$ ($R_{\odot}$) & \multicolumn{2}{c}{31.6 $^{+1.0}_{-0.6}$} & \multicolumn{2}{c}{30.4 $\pm$ 0.4}\\
$F_{\rm 2} / F_{\rm 1}$~(blue)        & \multicolumn{2}{c}{\phn0.70$^{+0.05}_{-0.01}$} & \multicolumn{2}{c}{\phs0.46 $\pm$ 0.02}\phn\\
\enddata
\tablenotetext{a}{Polar radius.}
\tablenotetext{b}{Radius toward the inner Lagrangian point.}
\end{deluxetable}

\clearpage

%
%

\input{epsf}

\clearpage

\input{epsf}

\clearpage

\input{epsf}

\clearpage

\input{epsf}

\clearpage

\input{epsf}

\clearpage

\input{epsf}

\end{document}